\documentclass[reprint,amsmath,superscriptaddress,amssymb,aps,twocolumn,prstab]{revtex4-1}
\usepackage{graphicx,subfigure,bm,amssymb,amsmath,dcolumn,hyperref}
\usepackage{color,multirow}
\usepackage{mathrsfs}
\usepackage{enumitem}

\begin{document}
\newcommand{\blue}[1]{\textcolor{blue}{#1}}
\newcommand{\red}[1]{\textcolor{red}{#1}}
\newcommand{\green}[1]{\textcolor{green}{#1}}
\newcommand{\orange}[1]{\textcolor{orange}{#1}}

\newcommand{\Lm}{L_{\rm min}}
\newcommand{\Vm}{V_{\rm min}}
\newcommand{\scrO}{\mathcal{O}}
\newcommand{\scrC}{\mathcal{C}}
\newcommand{\scrL}{\mathcal{L}}
\newcommand{\scrH}{\mathcal{H}}
\newcommand{\scrZ}{\mathcal{Z}}
\newcommand{\scrG}{\mathcal{G}}
\newcommand{\scrV}{\mathcal{V}}
\newcommand{\scrE}{\mathcal{E}}
\newcommand{\scrA}{\mathcal{A}}
\newcommand{\scrF}{\mathcal{F}}
\newcommand{\scrB}{\mathcal{B}}
\newcommand{\Clone}{F_1}
\newcommand{\Cltwo}{F_2}

\newcommand{\ZZ}{\mathbb{Z}}
\newcommand{\EE}{\mathbb{E}}

\newcommand{\yt}{y_t}
\newcommand{\yh}{y_h}

\title{Logarithmic Finite-Size Scaling of the Four-Dimensional Ising Model }
\author{Zhiyi Li}
\thanks{These two authors contributed equally to this paper.}
\affiliation{Department of Modern Physics, University of Science and Technology of China, Hefei, Anhui 230026, China}	

\author{Tianning Xiao}
\thanks{These two authors contributed equally to this paper.}
\affiliation{Hefei National Research Center for Physical Sciences at the Microscale,
University of Science and Technology of China, Hefei 230026, China}
\author{Zongzheng Zhou}
\email{eric.zhou@monash.edu}
\affiliation{School of Mathematics, Monash University, Clayton, Victoria 3800, Australia}
\author{Sheng Fang}
\email{11132024012@bnu.edu.cn}
\affiliation{School of Systems Science and Institute of Nonequilibrium Systems, Beijing Normal University, Beijing 100875, China}
\affiliation{Hefei National Research Center for Physical Sciences at the Microscale,
University of Science and Technology of China, Hefei 230026, China}

\author{Youjin Deng}
\email{yjdeng@ustc.edu.cn}
\affiliation{Department of Modern Physics, University of Science and Technology of China, Hefei, Anhui 230026, China}
\affiliation{Hefei National Research Center for Physical Sciences at the Microscale,
University of Science and Technology of China, Hefei 230026, China}
\affiliation{Hefei National Laboratory, University of Science and Technology of China, Hefei 230088, China}

\begin{abstract}
Field-theoretical calculations predict that, at the upper critical dimension $d_c=4$, the finite-size scaling (FSS) behaviors of the Ising model would be modified by multiplicative logarithmic corrections with thermal and magnetic correction exponents $(\hat{y}_t, \hat{y}_h)=(1/6,1/4)$. Using high-efficient cluster algorithms and the lifted worm algorithm, we present a systematic study to the FSS of the four-dimensional Ising model at criticality in the Fortuin-Kasteleyn (FK) bond and loop representations. In the FK representation, the size of the largest cluster is observed to scale as $C_1\sim L^3 (\ln L)^{\hat{y}_h}$, while the size of the second-largest cluster scales as $C_2\sim L^3 (\ln L)^{\hat{y}_{h2}}$ with $\hat{y}_{h2} = -1/4$ a new correction exponent not yet predicted from field theory. In the loop representation, we observe that the size of the largest loop cluster scales as $F_1 \sim L^{2} (\ln L)^{\hat{y}_t}$, and the specific heat scales as $c_{\textsc{e}} \sim (\ln L)^{2\hat{y}_t}$. This clarifies the long-standing open question that whether the specific heat for the critical Ising model at $d_c  = 4$ diverges logarithmically.

\end{abstract}
\date{\today}
\maketitle
\section{introduction}
The Ising model is one of the most fundamental models in statistical physics and plays a crucial role in facilitating the comprehensive analysis of phase transitions and critical phenomena~\cite{duminil2022100}. For a lattice $G=(V,E)$ with the vertex set $V$ and edge set $E$, the Hamiltonian of the Ising model reads
\begin{equation}
\label{eq:model_spin}
\scrH(s) = - J\sum_{ij \in E} s_i\cdot s_j\ + h\sum_i s_i \;, 
\end{equation}
where  $s_i\in\{-1,+1\}$ denotes the spin on the $i$-th vertex and the summation runs over all edges on the lattice, $h$ refers to the external magnetic field and $J>0$ is the coupling strength. The partition function is then given by $\mathcal{Z} = \sum_{s} e^{-\beta\scrH(s)}$. Let $K := \beta J$ be the reduced coupling strength. Hereinafter, we set $J = 1$ and focus on the zero-field case with $h = 0$ in this paper. If the spin is extended to be an $n$-component vector with unit length, then Eq.~\eqref{eq:model_spin} is the Hamiltonian of the O($n$) model~\cite{ONmodel}, where the case $n=1$ is the Ising model.

In most cases, the Ising model cannot be exactly solved, making the investigation of its critical behaviors heavily reliant on numerical methods on finite systems, such as Monte Carlo (MC) simulations. To analyze the results, the finite-size scaling (FSS) method is employed as a powerful method, which describes the asymptotic approach of finite systems to the thermodynamic limit near a continuous phase transition point $K_c$, to effectively estimate critical points and exponents~\cite{SuzukiFSS,FSSPriv,brezin1985finite,nishimori2011elements}. 
The main assumption of FSS is that the correlation length is effectively truncated by the linear system size $L$, such that the singular part of the free energy density function for a $d$-dimensional system can be written as
\begin{equation}
    f(t,h) = L^{-d}\Tilde{f}(tL^{y_t},hL^{y_h}),
    \label{standardFSS}
\end{equation}
where $t= (K_c-K)/K_c$ measures the distance from the critical point, $y_t,y_h$ are the corresponding thermal and magnetic renormalization group (RG) exponents and $\Tilde{f}(\cdot)$ is a scaling function. The FSS behaviors of various macroscopic quantities can be derived through the free energy function accordingly. For example, the magnetic susceptibility $\chi$ and the specific heat $c_{\textsc{e}}$ at the critical point and without the external field scale as, 
\begin{align}
    c_{\textsc{e}} & = -\frac{\partial^2 f}{\partial t^2} \sim L^{2y_t-d}, \label{cv_FSS}\\
    \chi & = -\frac{\partial^2 f}{\partial h^2} \sim L^{2y_h-d}. \label{chi_FSS}
\end{align}
Besides, the FSS theory also hypothesizes that at the critical point, the spin-spin correlation function $g(\mathbf{r},L) = \langle s_0 s_{\mathbf{r}}\rangle$ decays with distance $r$ as 
\begin{equation}
    g(\mathbf{r},L) \asymp \|\mathbf{r}\|^{-2(d-y_h)}\Tilde{g}(\|\mathbf{r}\|/L),
\end{equation}
where $\Tilde{g}(\cdot)$ is a scaling function.
\par
Above the upper critical dimension, $d > d_c = 4$, the scaling behaviors of the Ising model are characterized by the RG exponents given by mean field theory \cite{PhysRevLett.47.1}, or, specifically, by the Gaussian fixed point (GFP) as $(y_t,y_h) =(2,1+d/2)$. However, if one considers the system with periodic boundary condition (PBC), such standard FSS breakdowns \cite{brezin1985finite,BDofFSS,FSSforhighDIsingtori}.  For example, it is observed that $\chi \sim L^{d/2}$ in the PBC case, which is different from $\chi \sim L^2$ as the standard FSS in Eq.~\eqref{chi_FSS} predicts. Recent numerical and theoretical results suggest that the scaling form of the free energy function, as described by Eq.~\eqref{standardFSS}, is conjectured to have an extended form as \cite{5DFK2020}
\begin{equation}
    f(t,h) = L^{-d}\Tilde{f}_0(tL^{y_t},hL^{y_h})+L^{-d}\Tilde{f}_1(tL^{y_t^*},hL^{y_h^*}),
    \label{highDFSS}
\end{equation}
where $(y_t^*,y_h^*)=(d/2,3d/4)$ are obtained from the exact calculation of the complete-graph (CG) Ising model~\footnote{A complete graph with V vertices is a graph on which each vertex is connected to all others}, which can be seen as the application of the Landau mean field theory to finite systems~\cite{luijten1997interaction}. The $\Tilde{f}_0$ term corresponds to the GFP asymptotics, accounts for spatial fluctuations and governs the FSS of distance-dependent observables. The $\Tilde{f}_1$ term represents the CG asymptotics and governs the leading FSS of various macroscopic observables, such as $\chi$ and $c_{\textsc{e}}$. 

\par
At $d_c = 4$, the two sets of mean-field RG exponents coincide with each other, i.e., $(y_t, y_h) = (y^*_t, y^*_h)$. Field theory predicts that at the upper critical dimension, multiplicative logarithmic corrections appear. For the O$(n)$ model, in the thermodynamic limit, when approaching the criticality $(t\to0)$, susceptibility and specific heat are predicted to diverge as~\cite{larkin1969phase,RGanalysis1973}

\begin{align}
\chi(t) \sim& |t|^{-1}(-\ln|t|)^{\frac{n+2}{n+8}},  \\  
c_{\textsc{e}}(t) \sim& (-\ln|t|)^{\frac{4-n}{n+8}}.
\end{align}
In terms of the FSS, the singular part of the finite-size free-energy density involving the multiplicative logarithmic corrections for the O$(n)$ model is proposed  as
\begin{align}
    f(t,h) = & L^{-4}\Tilde{f}(tL^{y_t}(\ln{L})^{\hat{y}_t},hL^{y_h}(\ln{L})^{\hat{y}_h}),
    \label{log-correction-Kenna}
\end{align}
with $\hat{y}_t = (4-n)/(2n+16),\hat{y}_h = 1/4$~\cite{kenna2004finite,Kenna2006Logarithmic,kennaSCscaling}. The $n=1$ case, which is the Ising model, is studied in Ref.~\cite{aktekin2001finite} where $(\hat{y}_t,\hat{y}_h)=(1/6,1/4)$. Recently, as inspired from the high-dimensional scaling form (Eq.~\eqref{highDFSS}), the FSS of the free energy density at $d_c = 4$ is conjectured in Ref.~\cite{lv2021finite} as
\begin{align}
    f(t,h) =& L^{-4}\Tilde{f}_0(tL^{y_t},hL^{y_h}) \notag\\
    &+L^{-4}\Tilde{f}_1(tL^{y_t}(\ln{L})^{\hat{y}_t},hL^{y_h}(\ln{L})^{\hat{y}_h}),
    \label{log-correction}
\end{align}
and the correlation function is conjectured as 
\begin{equation}
    g(\mathbf{r},L) \sim 
    \left\{ \begin{aligned}
    & \|\mathbf{r}\|^{-2}, & \ \ \|\mathbf{r}\| \leq O(L/(\ln L)^{2\hat{y}_h})& \\
    & L^{-2}(\ln L)^{2\hat{y}_h}, &  \ \ \|\mathbf{r}\| > O(L/(\ln L)^{2\hat{y}_h})&
    \end{aligned} \right. .
    \label{correlation4D}
\end{equation}
Compared with Eq.~\eqref{log-correction-Kenna}, the key feature in Eq.~\eqref{log-correction}-\eqref{correlation4D} is the simultaneous existence of the GFP and the CG (after modified with multiplicative logarithmic corrections) asymptotics in the FSS formula of the free energy density, and the logarithmic corrections only apply to the CG term, not to the GFP term. Accordingly, the leading FSS of various macroscopic quantities suffer from logarithmic corrections, while quantities purely controlled by the GFP, such as the short-distance decay of $g(\mathbf{r},L)$, are free from logarithmic corrections.
Specifically, for the 4D Ising model at the critical point, from Eqs.~\eqref{log-correction} and~\eqref{correlation4D}, one can obtain the FSS behaviors for the following quantities:
 \begin{enumerate}[label=(\roman*)]
     \item The specific heat $c_{\textsc{e}} \sim (\ln{L})^{2\hat{y}_t}$ ;
     \item The magnetic susceptibility $\chi \sim L^2(\ln{L})^{2\hat{y}_h}$;
     \item The magnetic fluctuations at $\mathbf{k} \neq 0$ Fourier modes $\chi_{\mathbf{k}} \sim L^2$, since it is purely determined from the short-distance behaviour of $g(\mathbf{r},L)$. Here $\chi_{\mathbf{k}}$ is defined as $\chi_{\mathbf{k}} \equiv L^{-d} \langle |\mathcal{M}_{\mathbf{k}}| \rangle $, where $\mathcal{M}_{\mathbf{k}} \equiv \sum_{\mathbf{r}} s_\mathbf{r} e^{i\mathbf{k}\cdot \mathbf{r}}$ represents the Fourier mode of magnetization.
 \end{enumerate}

However, numerically verifying the logarithmic corrections is a challenging task. Earlier in 1987, a logarithmic scaling behavior of the specific heat was roughly observed in the 4D Ising model from a Monte Carlo (MC) study in Ref.~\cite{ESanchez-Velasco_1987}. However, since the simulated system size was not large enough, the result was inconclusive. Recently, a systematic numerical study in Ref.~\cite{lv2021finite} has been done to verify the multiplicative logarithmic corrections described by Eqs.~\eqref{log-correction} and~\eqref{correlation4D} in the 4D Ising, XY and Heisenberg models, respectively corresponding to $n=1,2,3$ case of the O$(n)$ model. The FSS behaviors of $\chi$, $\chi_\mathbf{k}$ and $g(\mathbf{r}, L)$ are observed to be consistent with the predictions from Eqs.~\eqref{log-correction} and~\eqref{correlation4D}, providing strong evidence to the existence of the magnetic correction exponent $\hat{y}_h$. However, the expected logarithmic divergence of the specific heat has not been clearly observed in Ref.~\cite{lv2021finite}.
Later on, in Ref.~\cite{4DSAW}, the authors numerically studied the logarithmic FSS for the 4D self-avoiding walk(SAW) model, which corresponds to the $n\rightarrow 0$ case of the O$(n)$ model, and both the exponents $\hat{y}_t= \hat{y}_h=1/4$ were clearly observed. So, it turns out that observing the thermal correction exponent $\hat{y}_t$ for the Ising model is much more challenging. 
In fact, in Refs.~\cite{lundow2023revising,lundow2009Ising}, the authors suggest that the specific heat of the Ising model is bounded, i.e., $\hat{y}_t = 0$. Meanwhile, a large-size simulation up to $L=1024$ in Ref.~\cite{HOTRGfor4DIsing} using higher-order tensor RG method also fails to detect the multiplicative logarithmic correction in the specific heat.

Besides the conventional spin representation, the Ising model can also be effectively described and analyzed using two geometric representations: the Fortuin-Kasteleyn (FK) bond representation and the loop representation \cite{fortuin1972random,van1941lange}, which will be called the FK Ising model and the loop Ising model. These representations are derived from certain expansions of the partition function, offering alternative perspectives for studying the properties of the Ising model. Specifically, the FK Ising model is the $q = 2$ case of the general $q$-state random-cluster (RC) model. Given a graph $\mathcal{G}$, each edge of the graph is either occupied by a bond or not. Then, the partition function of the $q$-state RC model is
\begin{equation}
    \mathcal{Z}_{\rm FK} = \sum_{A \subset \mathcal{G}} q^{k(A)}v^{|A|},
\end{equation}
where $\sum_{A \subset \mathcal{G}}$ sums over all bond configurations, $v$ is the statistical weight of each occupied bond, $k(A)$ is the number of connected clusters on $A$ and $|A|$ denotes the number of bonds on $A$. For the $q=2$ case, the bond weight $v=e^{2K}-1$, where $K$ is the reduced coupling strength mentioned before.

Similarly, the loop Ising model is characterized by occupied bonds on edges. Yet, the configurations within this representation are uniquely restricted to Eulerian graphs, also known as even graphs, in which the number of incident occupied bonds for any vertex is even. If one denotes ${\rm even}( G)$ as the set of even subgraphs on $G$, the partition function of the loop Ising model is given by 
\begin{equation}
    \mathcal{Z}_{\rm Loop} = \sum_{ F \subseteq  G} w^{|F|} \delta_{ F \in {\rm even}( G)},
    \label{loop-partition}
\end{equation}
where $w = \tanh{K}$ is the weight of each occupied bond and the Kronecker delta function $\delta_{F \in {\rm even}( G)}$ serves as an indicator function that ensures that for any subgraph $F$ giving a nonzero contribution to the partition function is an even graph. Such representation is also called the random-current model, random even graph or the flow representation of the Ising model~\cite{HD2016RC,GrimmiteRandomEvenGraph}.

Recently, many numerical studies in Refs.~\cite{HighDFK1,HighDFK2,5DFK2020,FKCG2021, LoopCG2023,  xiao2023finitesize} have been conducted on the two geometric representations of the Ising model above $d_c = 4$ and on the CG (which can be regarded as the $d\to\infty$ limit). Compared to the spin Ising model, the FK and loop Ising models exhibit much richer geometric properties, such as two length scales, two configuration sectors and two scaling windows. In particular, numerical results strongly suggest there simultaneously exist two upper critical dimensions ($d_c = 4, d_p = 6$) in the FK Ising model; one can refer to Fig.~17 in Ref.~\cite{HighDFK1} and Tabel~I in Ref~\cite{xiao2023finitesize} for the summary of main results. Generally speaking, the scaling behavior of the largest FK cluster and large loop clusters (size $\gg O(L^2)$) are controlled by the CG Ising asymptotics, while other FK clusters and medium loop clusters (size $\leq O(L^2)$) are described by the GFP asymptotics. Moreover, a multiplicative logarithmic correction has been discovered in the scaling of the second-largest FK cluster as $C_2\sim L^{y_h}(\ln{L})^{-1}$ for all $d>4$, which has not been found by any theoretical investigation.

Inspired by these insightful results, we systematically explore the logarithmic corrections to various geometric quantities of the 4D FK and loop Ising models with PBC, in particular to check whether the thermal correction exponent $\hat{y}_t$ or the expected divergence of the specific heat can be clearly observed in geometric representations. We use Wolff and Swendsen-Wang algorithms to simulate the FK Ising model, and the lifted worm algorithm to simulate the loop Ising model. We study the scaling of the susceptibility $\chi$ in the FK and loop Ising models, and both suggest that $\chi \sim L^2(\ln{L})^{1/2}$, consistent to the theoretical prediction and previous numerical results in Ref.~\cite{lv2021finite}. Moreover, we also observe that the largest FK cluster $C_1$, which is a magnetic quantity, scales as $L^3(\ln{L})^{\hat{y}_h}$ with $\hat{y}_h = 1/4$. The sizes of loop clusters in the loop Ising model are energy-like quantities. Since it is conjectured in Eq.~\eqref{log-correction} that the logarithmic corrections only apply to the modified CG term and the large loop clusters are believed to follow the CG-loop-Ising asymptotics, we hope the large loop clusters suffer much less additive finite-size corrections from the GFP term compared with energy quantities in the spin Ising model. Indeed, our results suggest that the size of the largest loop cluster scales as $F_1 \sim L^{y_t}(\ln{L})^{\hat{y}_t}$ with $y_t = 2$ and $\hat{y}_t = 1/6$. In addition, we study the variance of the number of loop bonds $c_{\textsc{b}}$, which is shown to exhibit the same leading FSS as the spin specific heat, and our data strongly suggest $c_{\textsc{b}} \sim (\ln{L})^{2\hat{y}_t}$. Thus, the long-standing subtle question on the scaling of the specific heat of the spin Ising model is clarified in the loop Ising model.

Additionally, we study the size of the second-largest cluster $C_2$ in the FK Ising model and our data suggest that $C_2 \sim L^3(\ln{L})^{-1/4}$, implying a new magnetic logarithmic correction exponent $\hat{y}_{h2} = -1/4$, which has no theoretical prediction. Therefore, even in terms of logarithmic corrections, the Ising model under geometric representations also exhibit richer phenomena than the spin Ising model. 

\par
The remainder of this paper is organized as follows. Section \ref{secII} summarizes the details of the simulation and the samples. Section \ref{secIII} contains our main numerical results.  A discussion is given in Sec.~\ref{secIV}.

\begin{figure*}
    \centering
    \includegraphics[width = 0.8\textwidth]{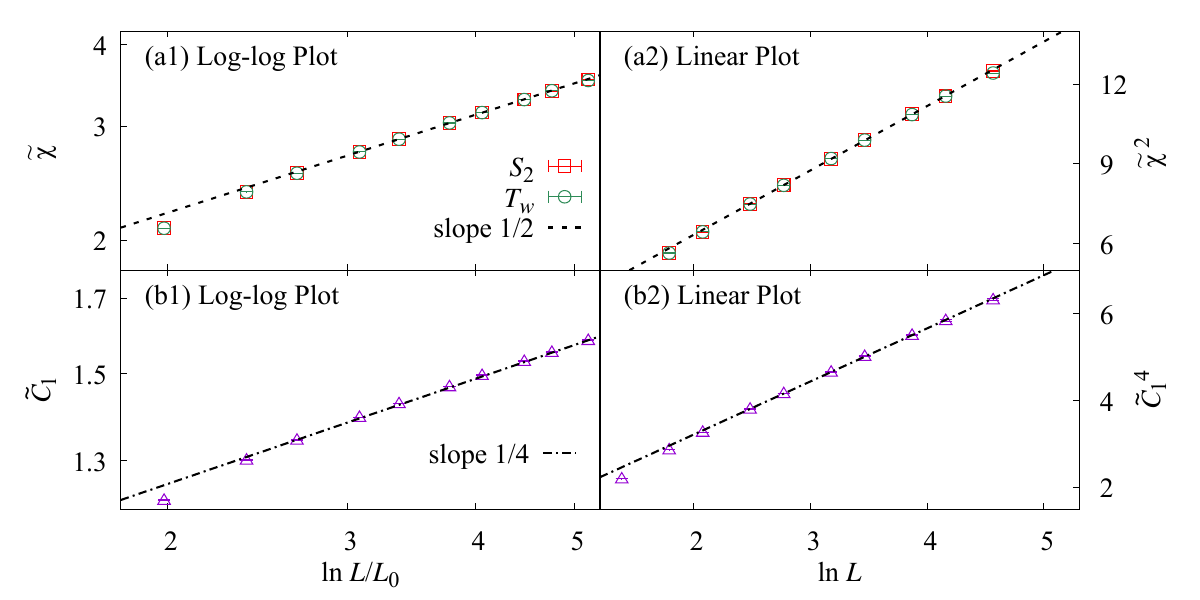}
    \caption{(a1) Log-log plot of the rescaled susceptibility $\Tilde{\chi}:=\chi/L^2$ in the two geometric representations, namely, the second moment of clusters' size $S_2$ in the FK representation and the worm returning time $T_w$ in the loop representation, against $\ln{L/L_0}$, with the constant $L_0$ fixed at $0.55$. The nice data collapse onto the dashed line with slope $1/2$ implies that $\chi \sim L^2(\ln{L})^{1/2}$, following the theoretical prediction. (a2) Plot of $\Tilde{\chi}^2$ versus $\ln{L}$ in the standard coordinate system. The obvious linearity further confirms the logarithmic FSS of $\chi$, but without the need to fix $L_0$. (b1) Log-log plot of the rescaled largest cluster $\Tilde{C}_1 := C_1/L^3$ versus $\ln{L/L_0}$ with $L_0=0.55$. The data clearly collapse onto a dashed line with a slope of $1/4$, implying the scaling $C_1 \sim L^3 (\ln L)^{1/4}$. (b2) Plot of $\Tilde{C}_1^4$ versus $\ln{L}$. The clear linear relation strongly supports the aforementioned logarithmic scaling of $C_1$.}
    \label{fig:FK-cluster}
\end{figure*}
\par
\section{simulation and observable}
\label{secII}

In our study, we have performed simulations of the four-dimensional Ising model with PBC employing a hybrid approach that combines the Swendsen-Wang (SW) algorithm \cite{SWalgorithm} with the Wolff \cite{WolffAlg} algorithm at the critical point $K_c=0.149\,693\,785(20)$~
\cite{lundow2023revising}. The SW algorithm is utilized to generate the FK cluster configurations, while the Wolff algorithm is applied between consecutive SW steps to update the spin configurations, since it is believed that the
Wolff algorithm has a smaller dynamic exponent than the SW
algorithm \cite{PhysRevE.66.057101}. In particular, we sample the following observables:
\begin{enumerate}[label=(\roman*)]
    \item The size of the largest cluster $\mathcal{C}_1$ and the second-largest cluster $\mathcal{C}_2$;
    \item The second moment of cluster size $\mathcal{S}_2 = (\sum_i \scrC_i^2)/L^4$, where $\scrC_i$ is the size of the $i$-th large cluster;
\end{enumerate}\par
Meanwhile, we employ the lifted worm algorithm\cite{WormAlg,LiftedWorm} to generate the loop configuration and sample the following observables in the loop representation
\begin{enumerate}[label=(\roman*)]
    \item The size of the largest loop cluster $\mathcal{F}_1$ and the second-largest loop cluster $\mathcal{F}_2$; 
    \item The total number of bonds $\mathcal{B}$ in loop clusters;
    \item The returning time of each worm update $\mathcal{T}_w$, namely, the Monte Carlo steps for generating a new loop configuration;
\end{enumerate}
By taking the ensemble average $\langle\cdot\rangle$ of these observables, we calculate the following quantities
\begin{enumerate}[label=(\roman*)]
    \item The mean size of the largest FK cluster $C_1= \langle \mathcal{C}_1\rangle$ and the second-largest FK cluster $C_2 = \langle \mathcal{C}_2 \rangle$;
    \item The second moment of cluster size in the FK representation $S_2 = \langle \mathcal{S}_2 \rangle$;
    \item The average number of bonds in the loop representation $B = \langle \scrB \rangle$, an energy-like quantity, and its variance $c_{\textsc{b}} = L^{-d}(\langle \mathcal{B}^2 \rangle - \langle\mathcal{B}\rangle^2)$, which we show is a linear function of the specific heat $c_{\textsc{e}}$;
    \item The average returning time $T_w = \langle \mathcal{T}_w\rangle$, which is equivalent to the spin susceptibility $\chi$ in the spin representation \cite{WormAlg};
    \item The mean size of the largest loop cluster $F_1= \langle \mathcal{F}_1\rangle$ and the second-largest loop cluster $F_2 = \langle \mathcal{F}_2 \rangle$.
\end{enumerate}
For both the FK Ising and the loop Ising models, the largest system size we simulate is $L_{\rm max} = 96$, containing about $10^8$ lattice sites. Approximately, for each system size, the number of independent samples are between $10^6$ and $4\times 10^6$. 

Here we employ a pseudorandom number generator based on the modulo-2 addition of two independent shift registers with lengths chosen as the Mersenne exponents 127 and 9689. This generator is well tested in Ref.~\cite{PseudoRN}, and no biased error has been found thus far.
\section{results}
\label{secIII}

\subsection{Magnetic scaling behaviors}

In this section, we first discuss the scaling behaviors of some magnetic quantities in geometric representations. It can be shown that both the second moment of sizes of all FK-clusters $S_2$ and the returning time $T_w$ correspond to the spin susceptibility $\chi$~\cite{PhysRevD.38.2009,grimmett2006random,PhysRevLett.87.160601}.

Since $y_h = 3$ in 4D, the leading power-law scaling of $\chi$ is expected to be $L^2$. Therefore, to see whether there are multiplicative corrections in the FSS of $\chi$, we study a rescaled susceptibility $\Tilde{\chi}:= \chi/L^2$.  In Fig.~\ref{fig:FK-cluster}(a1), we plot two sets of $\Tilde{\chi}$ from $S_2$ and $T_w$ respectively versus $\ln(L/L_0)$ in log-log scale, where $L_0$ is a non-universal constant and we fix $L_0 = 0.55$. One can see that, asymptotically, the data of $\Tilde{\chi}$ collapse onto the dashed line with slope $1/2$, suggesting the scaling $\chi \sim L^2 \left[\ln(L/L_0)\right]^{1/2}$. To remove the uncertainties caused by the constant $L_0$, we plot $\Tilde{\chi}^2$ against $\ln L$ in Fig.~\ref{fig:FK-cluster}(a2). The good data collapse onto the straight line suggests that $\chi \sim L^2 (a_1\ln L + a_2)^{1/2}$, with some constants $a_1$ and $a_2$. Thus it confirms the expected scaling $\chi \sim L^2 (\ln L)^{2\hat{y}_h}$ with $\hat{y}_h = 1/4$.

Next, we examine the effect of logarithmic corrections on the size of clusters in the FK Ising model. It is expected that the fractal dimension of the FK clusters is equal to the magnetic exponent $y_h$, as numerically observed above 4D in Refs.~\cite{HighDFK1,HighDFK2}. Therefore, one can expect that at 4D, the power-law scaling behavior of the FK clusters is dominated by $L^{y_h}$ with $y_h = 3$. Likewise, we introduce the rescaled cluster sizes $\Tilde{C}_n := C_n/L^3(n=1,2)$ for the largest and second-largest clusters $C_1$ and $C_2$. In Fig.~\ref{fig:FK-cluster}(b1), we plot in log-log scale $\Tilde{C}_1$ versus $\ln{L/L_0}$ with $L_0$ fixed at $0.55$. The data collapse nicely onto a straight line with slope $1/4$, indicating that $C_1 \sim L^{3} [\ln (L/L_0)]^{1/4}$. Similar to Fig.~\ref{fig:FK-cluster}(a2), we plot $\Tilde{C}_1^4$ against $\ln{L}$ in Fig.~\ref{fig:FK-cluster}(b2), to remove the uncertainty from the constant $L_0$. Again, the nice linear relationship implies that $C_1 \sim L^3({a_1\ln L + a_2})^{1/4}$. So at 4D, our data suggest that $C_1 \sim L^{y_h}({\ln L})^{\hat{y}_h}$ with $\hat{y}_h = 1/4$.

\begin{figure}
    \centering
    \includegraphics[width = 0.9\linewidth]{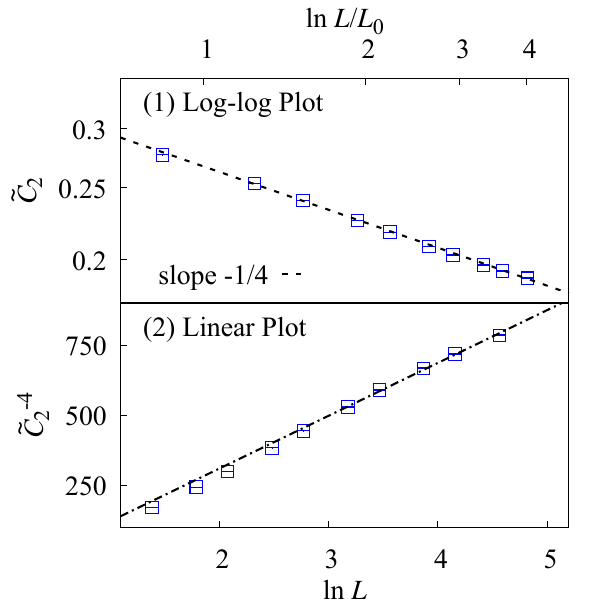}
    \caption{(top) Log-log plot of the second largest FK cluster $\Tilde{C}_2 := C_2/L^3$ versus $\ln{L/L_0}$ with $L_0 = 1.73$. The data collapse onto a dashed line with slope $-1/4$. (bottom) Plot of $\Tilde{C}_2^{-4}$ as a function of $\ln{L}$ in the standard scale, highlighted by a dashed straight line. These plots strongly suggest that the multiplicative logarithmic correction for the second-largest FK cluster is dominated by $(\ln L)^{-1/4}$.}
    \label{fig:C2-scaling}
\end{figure}

Then we investigate the scaling behavior of the second-largest cluster $C_2$.  In the high-d FK Ising model ($d>4$), a multiplicative logarithmic correction has been observed for the second-largest cluster, which scales as $C_2 \sim L^{1+d/2}({\ln L})^{-1}$~\cite{HighDFK1}. This indicates at 4D, the second-largest cluster may exhibit different logarithmic corrections to the largest cluster. Figure~\ref{fig:C2-scaling}(1) plots $\Tilde{C}_2$ versus $\ln{L/L_0}$ in the log-log scale, with $L_0$ fixed at $1.73$. As one can see, the correction term decreases as the system size increases, which is totally different from $C_1$ shown in Fig.~\ref{fig:FK-cluster}(b1). The data collapse onto the dashed line with slope $-1/4$ in the figure suggests that $C_2 \sim L^3 (\ln{L})^{-1/4}$. Furthermore, we plot $1/\Tilde{C}_2^{4}$ versus $\ln{L}$ in Fig.~\ref{fig:C2-scaling}(2), and the data points collapse well onto a straight line. This confirms the aforementioned scaling behaviour for $C_2$, i.e., the logarithmic correction in $C_2$ is characterized by a new exponent $\hat{y}_{h2} = -1/4$. This finding is particularly noteworthy as the logarithmic correction in $C_2$ that has not been previously identified in either theoretical or numerical studies at $d_c = 4$. We note that this result is also different from the high-d cases ($d>d_c$), where $\hat{y}_{h2} = -1$.

\begin{figure*}[ht]
\centering
\includegraphics[width = 0.8\textwidth]{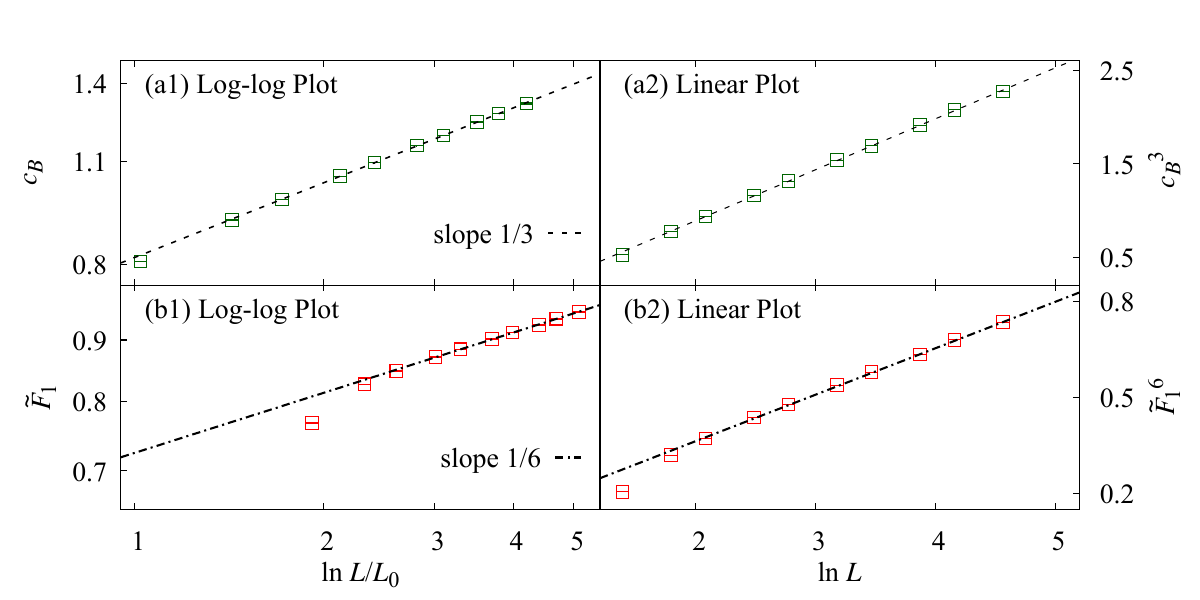}
\caption{Plots to demonstrate that the variance of the number of loop bonds scales as $c_{\textsc{b}} \sim (\ln L)^{2\hat{y}_t}$, and the size of the largest loop cluster scales as $F_1 \sim L^2 (\ln L)^{\hat{y}_t}$, where $\hat{y}_t = 1/6$. (a1) Log-log plot of $c_{\textsc{b}}$, which is proportional to the specific heat $c_{\textsc{e}}$, versus $\ln{L/L_0}$ with $L_0 = 1.44$. The data collapse onto a dashed line with slope $1/3$. (a2) Plot of $c_{\textsc{b}}^3$ versus $\ln{L}$. The linear relation further confirms the aforementioned scaling for $c_{\textsc{b}}$. (b1) Log-log plot of $\Tilde{F}_1 := F_1/L^2$ versus $\ln{L/L_0}$, with $L_0$ fixed at $0.59$. (b2) Plot of $\Tilde{F}_1^6$ versus $\ln{L}$. These two plots support the aforementioned scaling for $F_1$.}
\label{fig:Loop-Cluster}
\end{figure*}

\subsection{Thermal scaling behaviors}
In this section, we study the logarithmic corrections of some thermal quantities. From the partition function of the loop representation in Eq.~\eqref{loop-partition}, and the relation between the specific heat and the partition function, $c_{\textsc{e}} = K^2\frac{\partial ^2}{ \partial K^2} \ln{Z}$, we can derive at criticality that 
\begin{equation}
    c_{\textsc{e}} = \frac{K_c^2}{(\sinh K_c \cosh K_c)^{2}}c_{\textsc{b}} + a + O(L^{-d}) ,
    \label{specific heat}
\end{equation}
where $c_{\textsc{b}}$ is the variance of the number of bonds and $a = -4K_c^2\tanh^2{K_c}$ is a constant term. In other words, asymptotically, in the loop Ising model, the specific heat $c_{\textsc{e}}$ is proportional to the variance of bond number $c_{\textsc{b}}$. It is predicted from Eq.~\eqref{log-correction} that at $d=4$, 
\begin{equation}
    c_{\textsc{e}} \sim a_0 + a_1\left[\ln{(L/L_0)}\right]^{2\hat{y}_t}\;,
\label{Eq:specific-heat-scaling}
\end{equation}
with $\hat{y}_t = 1/6$. To confirm this scaling numerically, we first plot in Fig.~\ref{fig:Loop-Cluster}(a1) the data of $c_{\textsc{b}}$ versus $\ln{L/L_0}$ in the log-log plot, with $L_0$ fixed at $1.44$. The good data collapse onto the dashed line with $1/3$ suggests that $c_{\textsc{b}} \sim (\ln{L/L_0})^{1/3}$. To avoid the effect of fixing $L_0$ to some constant, we plot $c^3_{\textsc{e}}$ versus $\ln{L}$ in Fig.~\ref{fig:Loop-Cluster}(a2). The nice linear relationship provides a strong evidence to support the scaling that $c_{\textsc{b}} \sim (\ln{L})^{1/3}$. Combining with Eq.~\eqref{specific heat}, our data strongly suggest that at the critical 4D Ising model, the specific heat diverges as in Eq.~\eqref{Eq:specific-heat-scaling}. The difficulty in the previous analysis of the specific heat in the spin representation might arise from the effects of the background term $a_0$ in Eq.~\eqref{Eq:specific-heat-scaling}. Thus, the long-standing open question about whether the specific heat diverges logarithmically at the 4D Ising model is clarified from the loop Ising model.

We next explore the FSS of the size of the largest and second-largest loop clusters $F_1, F_2$ in the loop Ising model, in particular, to examine the form of logarithmic corrections. Different from the FK clusters, size of loop clusters are thermal quantities, with the fractal dimension equal to the thermal exponent $y_t$, as numerically observed for $d > 4$ in Ref.~\cite{xiao2023finitesize}. Thus at $d=4$, we expect $F_1,F_2 \sim L^2$, accompanied by multiplicative logarithmic corrections. To find the form of logarithmic corrections, again we plot the rescaled largest loop cluster size $\Tilde{F}_1 := F_1/L^2$ versus $\ln{L/L_0}$ in Fig.~\ref{fig:Loop-Cluster}(b1) with $L_0$ fixed at $0.59$. Except for the small systems which suffer strong finite-size effect, data with large system collapse nicely on the dashed line with slope $1/6$. Thus, asymptotically, our data support the scaling $F_1 \sim L^2 (\ln{L/L_0})^{1/6}$. To further confirm, we also plot in Fig.~\ref{fig:Loop-Cluster}(b2) the $\Tilde{F}_1^6$ versus $\ln L$; in this way, the uncertainty from fixing the constant $L_0$ is removed. The clear linear relation demonstrates that $F_1 \sim L^2 (a_1\ln L + a_2)^{1/6}$. Thus, our data strongly suggest that $F_1 \sim L^{y_t} (\ln L)^{\hat{y}_t}$ with $\hat{y}_t = 1/6$. We note that, in comparison with the specific heat $c_E$, the background term $a_0$ in $F_1 \sim  a_0 + a_1 L^2 [\ln (L/L_0)]^{1/6}$ plays a less important role. 

For the size of the second-largest loop cluster $F_2$, by the same analysis, our data show that $F_2$ exhibits the same power-law scaling as $F_1$, i.e. $F_2 \sim L^{2}$, but the effect of the logarithmic corrections is too weak to be numerically detected.

\section{Conclusion}
\label{secIV}
In this work, we carry out a systematic study on the logarithmic corrections in the finite-size scaling (FSS) of the four-dimensional (4D) Ising model under two geometric representations, i.e., the Fortuin-Kasteleyn (FK) random-cluster and the loop representations. The study of the 4D Ising model is significant due to its connection to the O$(n)$ model, the scalar sector of the standard model, and its relevance in various condensed matter systems, where logarithmic correction behaviors are expected to manifest at the three-dimensional quantum critical point (QCP)~\cite{zinn2021quantum,Higgs1964,higgsmodeinCMP}.

In both the FK Ising and loop Ising models, our data indicate that the finite-size scaling (FSS) of the critical susceptibility aligns with the expected scaling \(\chi \sim L^{2y_h - d}(\ln L)^{2\hat{y}_h}\), where \(y_h = 3\) and \(\hat{y}_h = 1/4\). This excellent consistency validates the effectiveness of our numerical simulation, suggesting that the deviation between the coupling we simulated at and the genuine critical coupling has a negligible impact on our results. Moreover, in the FK Ising model, our data suggest that the size of the largest FK cluster scales as $C_1 \sim L^{y_h} (\ln L)^{\hat{y}_h}$, and the size of the second-largest cluster scales as $C_2 \sim L^{y_h} (\ln L)^{\hat{y}_{h2}}$ with $\hat{y}_{h2} = -1/4$ a new correction exponent which has no theoretical prediction. In the loop Ising model, the specific heat is observed to scale as $c_{\textsc{e}} \sim a_0 + a_1\left[\ln (L/L_0)\right]^{2\hat{y_t}}$ with $\hat{y_t} = 1/6$, which numerically confirm the logarithmic divergence of the critical specific heat of the 4D Ising model. For the size of the largest loop cluster, our data suggest $F_1 \sim L^{y_t} (\ln L)^{\hat{y}_t}$ with $\hat{y}_t = 1/6$.

We finally provide a simple explanation for why the logarithmic divergence of the specific heat is much easier to be numerically observed in the loop Ising model. From the conjectured FSS ansatz~\eqref{log-correction}, the predicted FSS for the specific heat can be written as in Eq.~\eqref{Eq:specific-heat-scaling}, in which the constant $a_0$ is the background term and the $a_1$-term corresponds to the modified CG term. If $a_0$ is comparable to $a_1$ and $L_0$ is unknown, then it is hard to numerically extract the value of the exponent $\hat{y_t}$ from the data. But in the loop Ising model, our data show that the background term in $c_{\textsc{b}}$ is very small, so the logarithmic correction can be easily extracted using Fig.~\ref{fig:Loop-Cluster}(a2). Once the logarithmic correction is determined for $c_{\textsc{b}}$, the scaling of the specific heat follows directly from Eq.~\eqref{specific heat}. More convincing evidence can be seen from the size of the largest loop cluster $F_1$. As proposed in Ref.~\cite{xiao2023finitesize}, the FSS of $F_1$ completely follows the CG asymptotics for $d > d_c$. If this picture also holds at $d = d_c$, then at $d_c$ it follows from Eq.~\eqref{log-correction} that $F_1 \sim a_0 + a_1 L^{y_t}\left[ \ln (L/L_0)\right]^{\hat{y}_t}$, without the effect from the GFP term. Thus, if we study the ratio $F_1/L^{y_t}$, then in comparison with the logarithmic term, the background $a_0$-term becomes subdominant with order $L^{-y_t}$. In $c_{\textsc{b}}$, the effect of the background to the logarithmic term is also subdominant but of order $(\ln L)^{-\hat{y}_t}$, much larger than $L^{-y_t}$. In other words, in $F_1$, the background term has much weaker effect to the estimate of the exponent $\hat{y}_t$, and thus the result from $F_1$ is more convincing.

\section*{acknowledgement}
This work has been supported by the National Natural Science Foundation of China (under Grant No. 12275263), the Innovation Program for Quantum Science and Technology (under grant No. 2021ZD0301900), the Natural Science Foundation of Fujian Province of China (under Grant No. 2023J02032). 

\bibliographystyle{apsrev4-2}
\bibliography{main}

\begin{thebibliography}{45}%
\makeatletter
\providecommand \@ifxundefined [1]{%
 \@ifx{#1\undefined}
}%
\providecommand \@ifnum [1]{%
 \ifnum #1\expandafter \@firstoftwo
 \else \expandafter \@secondoftwo
 \fi
}%
\providecommand \@ifx [1]{%
 \ifx #1\expandafter \@firstoftwo
 \else \expandafter \@secondoftwo
 \fi
}%
\providecommand \natexlab [1]{#1}%
\providecommand \enquote  [1]{``#1''}%
\providecommand \bibnamefont  [1]{#1}%
\providecommand \bibfnamefont [1]{#1}%
\providecommand \citenamefont [1]{#1}%
\providecommand \href@noop [0]{\@secondoftwo}%
\providecommand \href [0]{\begingroup \@sanitize@url \@href}%
\providecommand \@href[1]{\@@startlink{#1}\@@href}%
\providecommand \@@href[1]{\endgroup#1\@@endlink}%
\providecommand \@sanitize@url [0]{\catcode `\\12\catcode `\$12\catcode
  `\&12\catcode `\#12\catcode `\^12\catcode `\_12\catcode `\%12\relax}%
\providecommand \@@startlink[1]{}%
\providecommand \@@endlink[0]{}%
\providecommand \url  [0]{\begingroup\@sanitize@url \@url }%
\providecommand \@url [1]{\endgroup\@href {#1}{\urlprefix }}%
\providecommand \urlprefix  [0]{URL }%
\providecommand \Eprint [0]{\href }%
\providecommand \doibase [0]{https://doi.org/}%
\providecommand \selectlanguage [0]{\@gobble}%
\providecommand \bibinfo  [0]{\@secondoftwo}%
\providecommand \bibfield  [0]{\@secondoftwo}%
\providecommand \translation [1]{[#1]}%
\providecommand \BibitemOpen [0]{}%
\providecommand \bibitemStop [0]{}%
\providecommand \bibitemNoStop [0]{.\EOS\space}%
\providecommand \EOS [0]{\spacefactor3000\relax}%
\providecommand \BibitemShut  [1]{\csname bibitem#1\endcsname}%
\let\auto@bib@innerbib\@empty
\bibitem [{\citenamefont {Duminil-Copin}(2022)}]{duminil2022100}%
  \BibitemOpen
  \bibfield  {author} {\bibinfo {author} {\bibfnamefont {H.}~\bibnamefont
  {Duminil-Copin}},\ }\href {https://ems.press/books/standalone/273/5392}
  {\bibfield  {journal} {\bibinfo  {journal} {Int. Cong. Math.}\ }\textbf
  {\bibinfo {volume} {1}},\ \bibinfo {pages} {164} (\bibinfo {year}
  {2022})}\BibitemShut {NoStop}%
\bibitem [{\citenamefont {Stanley}(1968)}]{ONmodel}%
  \BibitemOpen
  \bibfield  {author} {\bibinfo {author} {\bibfnamefont {H.~E.}\ \bibnamefont
  {Stanley}},\ }\href {https://doi.org/10.1103/PhysRevLett.20.589} {\bibfield
  {journal} {\bibinfo  {journal} {Phys. Rev. Lett.}\ }\textbf {\bibinfo
  {volume} {20}},\ \bibinfo {pages} {589} (\bibinfo {year} {1968})}\BibitemShut
  {NoStop}%
\bibitem [{\citenamefont {Suzuki}(1977)}]{SuzukiFSS}%
  \BibitemOpen
  \bibfield  {author} {\bibinfo {author} {\bibfnamefont {M.}~\bibnamefont
  {Suzuki}},\ }\href {https://doi.org/10.1143/PTP.58.1142} {\bibfield
  {journal} {\bibinfo  {journal} {Prog. theor. phys.}\ }\textbf {\bibinfo
  {volume} {58}},\ \bibinfo {pages} {1142} (\bibinfo {year}
  {1977})}\BibitemShut {NoStop}%
\bibitem [{\citenamefont {Privman}(1990)}]{FSSPriv}%
  \BibitemOpen
  \bibfield  {author} {\bibinfo {author} {\bibfnamefont {V.}~\bibnamefont
  {Privman}},\ }\href {https://doi.org/10.1142/1011} {\emph {\bibinfo {title}
  {Finite Size Scaling and Numerical Simulation of Statistical Systems}}}\
  (\bibinfo  {publisher} {World Scientific},\ \bibinfo {year}
  {1990})\BibitemShut {NoStop}%
\bibitem [{\citenamefont {Br{\'e}zin}\ and\ \citenamefont
  {Zinn-Justin}(1985)}]{brezin1985finite}%
  \BibitemOpen
  \bibfield  {author} {\bibinfo {author} {\bibfnamefont {E.}~\bibnamefont
  {Br{\'e}zin}}\ and\ \bibinfo {author} {\bibfnamefont {J.}~\bibnamefont
  {Zinn-Justin}},\ }\href@noop {} {\bibfield  {journal} {\bibinfo  {journal}
  {Nucl. Phys. B}\ }\textbf {\bibinfo {volume} {257}},\ \bibinfo {pages} {867}
  (\bibinfo {year} {1985})}\BibitemShut {NoStop}%
\bibitem [{\citenamefont {Nishimori}\ and\ \citenamefont
  {Ortiz}(2011)}]{nishimori2011elements}%
  \BibitemOpen
  \bibfield  {author} {\bibinfo {author} {\bibfnamefont {H.}~\bibnamefont
  {Nishimori}}\ and\ \bibinfo {author} {\bibfnamefont {G.}~\bibnamefont
  {Ortiz}},\ }\href@noop {} {\emph {\bibinfo {title} {Elements of phase
  transitions and critical phenomena}}}\ (\bibinfo  {publisher} {Oxford
  university press},\ \bibinfo {year} {2011})\BibitemShut {NoStop}%
\bibitem [{\citenamefont {Aizenman}(1981)}]{PhysRevLett.47.1}%
  \BibitemOpen
  \bibfield  {author} {\bibinfo {author} {\bibfnamefont {M.}~\bibnamefont
  {Aizenman}},\ }\href {https://doi.org/10.1103/PhysRevLett.47.1} {\bibfield
  {journal} {\bibinfo  {journal} {Phys. Rev. Lett.}\ }\textbf {\bibinfo
  {volume} {47}},\ \bibinfo {pages} {1} (\bibinfo {year} {1981})}\BibitemShut
  {NoStop}%
\bibitem [{\citenamefont {Grimm}\ \emph {et~al.}(2017)\citenamefont {Grimm},
  \citenamefont {El\ifmmode~\mbox{\c{c}}\else \c{c}\fi{}i}, \citenamefont
  {Zhou}, \citenamefont {Garoni},\ and\ \citenamefont {Deng}}]{BDofFSS}%
  \BibitemOpen
  \bibfield  {author} {\bibinfo {author} {\bibfnamefont {J.}~\bibnamefont
  {Grimm}}, \bibinfo {author} {\bibfnamefont {E.~M.}\ \bibnamefont
  {El\ifmmode~\mbox{\c{c}}\else \c{c}\fi{}i}}, \bibinfo {author} {\bibfnamefont
  {Z.}~\bibnamefont {Zhou}}, \bibinfo {author} {\bibfnamefont {T.~M.}\
  \bibnamefont {Garoni}},\ and\ \bibinfo {author} {\bibfnamefont
  {Y.}~\bibnamefont {Deng}},\ }\href
  {https://doi.org/10.1103/PhysRevLett.118.115701} {\bibfield  {journal}
  {\bibinfo  {journal} {Phys. Rev. Lett.}\ }\textbf {\bibinfo {volume} {118}},\
  \bibinfo {pages} {115701} (\bibinfo {year} {2017})}\BibitemShut {NoStop}%
\bibitem [{\citenamefont {Zhou}\ \emph {et~al.}(2018)\citenamefont {Zhou},
  \citenamefont {Grimm}, \citenamefont {Fang}, \citenamefont {Deng},\ and\
  \citenamefont {Garoni}}]{FSSforhighDIsingtori}%
  \BibitemOpen
  \bibfield  {author} {\bibinfo {author} {\bibfnamefont {Z.}~\bibnamefont
  {Zhou}}, \bibinfo {author} {\bibfnamefont {J.}~\bibnamefont {Grimm}},
  \bibinfo {author} {\bibfnamefont {S.}~\bibnamefont {Fang}}, \bibinfo {author}
  {\bibfnamefont {Y.}~\bibnamefont {Deng}},\ and\ \bibinfo {author}
  {\bibfnamefont {T.~M.}\ \bibnamefont {Garoni}},\ }\href
  {https://doi.org/10.1103/PhysRevLett.121.185701} {\bibfield  {journal}
  {\bibinfo  {journal} {Phys. Rev. Lett.}\ }\textbf {\bibinfo {volume} {121}},\
  \bibinfo {pages} {185701} (\bibinfo {year} {2018})}\BibitemShut {NoStop}%
\bibitem [{\citenamefont {Fang}\ \emph {et~al.}(2020)\citenamefont {Fang},
  \citenamefont {Grimm}, \citenamefont {Zhou},\ and\ \citenamefont
  {Deng}}]{5DFK2020}%
  \BibitemOpen
  \bibfield  {author} {\bibinfo {author} {\bibfnamefont {S.}~\bibnamefont
  {Fang}}, \bibinfo {author} {\bibfnamefont {J.}~\bibnamefont {Grimm}},
  \bibinfo {author} {\bibfnamefont {Z.}~\bibnamefont {Zhou}},\ and\ \bibinfo
  {author} {\bibfnamefont {Y.}~\bibnamefont {Deng}},\ }\href
  {https://doi.org/10.1103/PhysRevE.102.022125} {\bibfield  {journal} {\bibinfo
   {journal} {Phys. Rev. E}\ }\textbf {\bibinfo {volume} {102}},\ \bibinfo
  {pages} {022125} (\bibinfo {year} {2020})}\BibitemShut {NoStop}%
\bibitem [{Note1()}]{Note1}%
  \BibitemOpen
  \bibinfo {note} {A complete graph with V vertices is a graph on which each
  vertex is connected to all others}\BibitemShut {NoStop}%
\bibitem [{\citenamefont {Luijten}(1997)}]{luijten1997interaction}%
  \BibitemOpen
  \bibfield  {author} {\bibinfo {author} {\bibfnamefont {E.}~\bibnamefont
  {Luijten}},\ }\href@noop {} {\emph {\bibinfo {title} {Interaction range,
  universality and the upper critical dimension}}}\ (\bibinfo  {publisher}
  {Delft University Press},\ \bibinfo {year} {1997})\BibitemShut {NoStop}%
\bibitem [{\citenamefont {Larkin}\ and\ \citenamefont
  {Khmel'Nitskiĭ}(1969)}]{larkin1969phase}%
  \BibitemOpen
  \bibfield  {author} {\bibinfo {author} {\bibfnamefont {A.~I.}\ \bibnamefont
  {Larkin}}\ and\ \bibinfo {author} {\bibfnamefont {D.~E.}\ \bibnamefont
  {Khmel'Nitskiĭ}},\ }\href
  {http://www.jetp.ras.ru/cgi-bin/dn/e_029_06_1123.pdf} {\bibfield  {journal}
  {\bibinfo  {journal} {Sov. Phys. JETP}\ }\textbf {\bibinfo {volume} {29}},\
  \bibinfo {pages} {1123} (\bibinfo {year} {1969})}\BibitemShut {NoStop}%
\bibitem [{\citenamefont {Wegner}\ and\ \citenamefont
  {Riedel}(1973)}]{RGanalysis1973}%
  \BibitemOpen
  \bibfield  {author} {\bibinfo {author} {\bibfnamefont {F.~J.}\ \bibnamefont
  {Wegner}}\ and\ \bibinfo {author} {\bibfnamefont {E.~K.}\ \bibnamefont
  {Riedel}},\ }\href {https://doi.org/10.1103/PhysRevB.7.248} {\bibfield
  {journal} {\bibinfo  {journal} {Phys. Rev. B}\ }\textbf {\bibinfo {volume}
  {7}},\ \bibinfo {pages} {248} (\bibinfo {year} {1973})}\BibitemShut {NoStop}%
\bibitem [{\citenamefont {Kenna}(2004)}]{kenna2004finite}%
  \BibitemOpen
  \bibfield  {author} {\bibinfo {author} {\bibfnamefont {R.}~\bibnamefont
  {Kenna}},\ }\href
  {https://doi.org/https://doi.org/10.1016/j.nuclphysb.2004.05.012} {\bibfield
  {journal} {\bibinfo  {journal} {Nucl. Phys. B}\ }\textbf {\bibinfo {volume}
  {691}},\ \bibinfo {pages} {292} (\bibinfo {year} {2004})}\BibitemShut
  {NoStop}%
\bibitem [{\citenamefont {Kenna}\ \emph
  {et~al.}(2006{\natexlab{a}})\citenamefont {Kenna}, \citenamefont {Johnston},\
  and\ \citenamefont {Janke}}]{Kenna2006Logarithmic}%
  \BibitemOpen
  \bibfield  {author} {\bibinfo {author} {\bibfnamefont {R.}~\bibnamefont
  {Kenna}}, \bibinfo {author} {\bibfnamefont {D.~A.}\ \bibnamefont
  {Johnston}},\ and\ \bibinfo {author} {\bibfnamefont {W.}~\bibnamefont
  {Janke}},\ }\href {https://doi.org/10.1103/PhysRevLett.96.115701} {\bibfield
  {journal} {\bibinfo  {journal} {Phys. Rev. Lett.}\ }\textbf {\bibinfo
  {volume} {96}},\ \bibinfo {pages} {115701} (\bibinfo {year}
  {2006}{\natexlab{a}})}\BibitemShut {NoStop}%
\bibitem [{\citenamefont {Kenna}\ \emph
  {et~al.}(2006{\natexlab{b}})\citenamefont {Kenna}, \citenamefont {Johnston},\
  and\ \citenamefont {Janke}}]{kennaSCscaling}%
  \BibitemOpen
  \bibfield  {author} {\bibinfo {author} {\bibfnamefont {R.}~\bibnamefont
  {Kenna}}, \bibinfo {author} {\bibfnamefont {D.~A.}\ \bibnamefont
  {Johnston}},\ and\ \bibinfo {author} {\bibfnamefont {W.}~\bibnamefont
  {Janke}},\ }\href {https://doi.org/10.1103/PhysRevLett.97.155702} {\bibfield
  {journal} {\bibinfo  {journal} {Phys. Rev. Lett.}\ }\textbf {\bibinfo
  {volume} {97}},\ \bibinfo {pages} {155702} (\bibinfo {year}
  {2006}{\natexlab{b}})}\BibitemShut {NoStop}%
\bibitem [{\citenamefont {Aktekin}(2001)}]{aktekin2001finite}%
  \BibitemOpen
  \bibfield  {author} {\bibinfo {author} {\bibfnamefont {N.}~\bibnamefont
  {Aktekin}},\ }\href {https://doi.org/https://doi.org/10.1023/A:1010457905088}
  {\bibfield  {journal} {\bibinfo  {journal} {J. Stat. Phys.}\ }\textbf
  {\bibinfo {volume} {104}},\ \bibinfo {pages} {1397} (\bibinfo {year}
  {2001})}\BibitemShut {NoStop}%
\bibitem [{\citenamefont {Lv}\ \emph {et~al.}(2021)\citenamefont {Lv},
  \citenamefont {Xu}, \citenamefont {Sun}, \citenamefont {Chen},\ and\
  \citenamefont {Deng}}]{lv2021finite}%
  \BibitemOpen
  \bibfield  {author} {\bibinfo {author} {\bibfnamefont {J.-P.}\ \bibnamefont
  {Lv}}, \bibinfo {author} {\bibfnamefont {W.}~\bibnamefont {Xu}}, \bibinfo
  {author} {\bibfnamefont {Y.}~\bibnamefont {Sun}}, \bibinfo {author}
  {\bibfnamefont {K.}~\bibnamefont {Chen}},\ and\ \bibinfo {author}
  {\bibfnamefont {Y.}~\bibnamefont {Deng}},\ }\href
  {https://doi.org/10.1093/nsr/nwaa212} {\bibfield  {journal} {\bibinfo
  {journal} {Natl. Sci. Rev.}\ }\textbf {\bibinfo {volume} {8}},\ \bibinfo
  {pages} {nwaa212} (\bibinfo {year} {2021})}\BibitemShut {NoStop}%
\bibitem [{\citenamefont {Sanchez-Velasco}(1987)}]{ESanchez-Velasco_1987}%
  \BibitemOpen
  \bibfield  {author} {\bibinfo {author} {\bibfnamefont {E.}~\bibnamefont
  {Sanchez-Velasco}},\ }\href {https://doi.org/10.1088/0305-4470/20/14/041}
  {\bibfield  {journal} {\bibinfo  {journal} {Journal of Physics A:
  Mathematical and General}\ }\textbf {\bibinfo {volume} {20}},\ \bibinfo
  {pages} {5033} (\bibinfo {year} {1987})}\BibitemShut {NoStop}%
\bibitem [{\citenamefont {Fang}\ \emph
  {et~al.}(2021{\natexlab{a}})\citenamefont {Fang}, \citenamefont {Deng},\ and\
  \citenamefont {Zhou}}]{4DSAW}%
  \BibitemOpen
  \bibfield  {author} {\bibinfo {author} {\bibfnamefont {S.}~\bibnamefont
  {Fang}}, \bibinfo {author} {\bibfnamefont {Y.}~\bibnamefont {Deng}},\ and\
  \bibinfo {author} {\bibfnamefont {Z.}~\bibnamefont {Zhou}},\ }\href
  {https://doi.org/10.1103/PhysRevE.104.064108} {\bibfield  {journal} {\bibinfo
   {journal} {Phys. Rev. E}\ }\textbf {\bibinfo {volume} {104}},\ \bibinfo
  {pages} {064108} (\bibinfo {year} {2021}{\natexlab{a}})}\BibitemShut
  {NoStop}%
\bibitem [{\citenamefont {Lundow}\ and\ \citenamefont
  {Markstr{\"o}m}(2023)}]{lundow2023revising}%
  \BibitemOpen
  \bibfield  {author} {\bibinfo {author} {\bibfnamefont {P.}~\bibnamefont
  {Lundow}}\ and\ \bibinfo {author} {\bibfnamefont {K.}~\bibnamefont
  {Markstr{\"o}m}},\ }\href
  {https://doi.org/https://doi.org/10.1016/j.nuclphysb.2023.116256} {\bibfield
  {journal} {\bibinfo  {journal} {Nucl. Phys. B}\ ,\ \bibinfo {pages} {116256}}
  (\bibinfo {year} {2023})}\BibitemShut {NoStop}%
\bibitem [{\citenamefont {Lundow}\ and\ \citenamefont
  {Markstr\"om}(2009)}]{lundow2009Ising}%
  \BibitemOpen
  \bibfield  {author} {\bibinfo {author} {\bibfnamefont {P.~H.}\ \bibnamefont
  {Lundow}}\ and\ \bibinfo {author} {\bibfnamefont {K.}~\bibnamefont
  {Markstr\"om}},\ }\href {https://doi.org/10.1103/PhysRevE.80.031104}
  {\bibfield  {journal} {\bibinfo  {journal} {Phys. Rev. E}\ }\textbf {\bibinfo
  {volume} {80}},\ \bibinfo {pages} {031104} (\bibinfo {year}
  {2009})}\BibitemShut {NoStop}%
\bibitem [{\citenamefont {Akiyama}\ \emph {et~al.}(2019)\citenamefont
  {Akiyama}, \citenamefont {Kuramashi}, \citenamefont {Yamashita},\ and\
  \citenamefont {Yoshimura}}]{HOTRGfor4DIsing}%
  \BibitemOpen
  \bibfield  {author} {\bibinfo {author} {\bibfnamefont {S.}~\bibnamefont
  {Akiyama}}, \bibinfo {author} {\bibfnamefont {Y.}~\bibnamefont {Kuramashi}},
  \bibinfo {author} {\bibfnamefont {T.}~\bibnamefont {Yamashita}},\ and\
  \bibinfo {author} {\bibfnamefont {Y.}~\bibnamefont {Yoshimura}},\ }\href
  {https://doi.org/10.1103/PhysRevD.100.054510} {\bibfield  {journal} {\bibinfo
   {journal} {Phys. Rev. D}\ }\textbf {\bibinfo {volume} {100}},\ \bibinfo
  {pages} {054510} (\bibinfo {year} {2019})}\BibitemShut {NoStop}%
\bibitem [{\citenamefont {Fortuin}\ and\ \citenamefont
  {Kasteleyn}(1972)}]{fortuin1972random}%
  \BibitemOpen
  \bibfield  {author} {\bibinfo {author} {\bibfnamefont {C.~M.}\ \bibnamefont
  {Fortuin}}\ and\ \bibinfo {author} {\bibfnamefont {P.~W.}\ \bibnamefont
  {Kasteleyn}},\ }\href@noop {} {\bibfield  {journal} {\bibinfo  {journal}
  {Physica}\ }\textbf {\bibinfo {volume} {57}},\ \bibinfo {pages} {536}
  (\bibinfo {year} {1972})}\BibitemShut {NoStop}%
\bibitem [{\citenamefont {van~der Waerden}(1941)}]{van1941lange}%
  \BibitemOpen
  \bibfield  {author} {\bibinfo {author} {\bibfnamefont {B.~L.}\ \bibnamefont
  {van~der Waerden}},\ }\href@noop {} {\bibfield  {journal} {\bibinfo
  {journal} {Zeitschrift f{\"u}r Physik}\ }\textbf {\bibinfo {volume} {118}},\
  \bibinfo {pages} {473} (\bibinfo {year} {1941})}\BibitemShut {NoStop}%
\bibitem [{\citenamefont {Duminil-Copin}(2018)}]{HD2016RC}%
  \BibitemOpen
  \bibfield  {author} {\bibinfo {author} {\bibfnamefont {H.}~\bibnamefont
  {Duminil-Copin}},\ }\href@noop {} {\bibfield  {journal} {\bibinfo  {journal}
  {European Congress of Mathematics, Eur. Math. Soc., Z\"urich}\ ,\ \bibinfo
  {pages} {869}} (\bibinfo {year} {2018})}\BibitemShut {NoStop}%
\bibitem [{\citenamefont {Grimmett}\ and\ \citenamefont
  {Janson}(2007)}]{GrimmiteRandomEvenGraph}%
  \BibitemOpen
  \bibfield  {author} {\bibinfo {author} {\bibfnamefont {G.}~\bibnamefont
  {Grimmett}}\ and\ \bibinfo {author} {\bibfnamefont {S.}~\bibnamefont
  {Janson}},\ }\href@noop {} {\bibfield  {journal} {\bibinfo  {journal} {The
  electronic journal of combinatoric}\ }\textbf {\bibinfo {volume} {16}},\
  \bibinfo {pages} {R46} (\bibinfo {year} {2007})}\BibitemShut {NoStop}%
\bibitem [{\citenamefont {Fang}\ \emph {et~al.}(2023)\citenamefont {Fang},
  \citenamefont {Zhou},\ and\ \citenamefont {Deng}}]{HighDFK1}%
  \BibitemOpen
  \bibfield  {author} {\bibinfo {author} {\bibfnamefont {S.}~\bibnamefont
  {Fang}}, \bibinfo {author} {\bibfnamefont {Z.}~\bibnamefont {Zhou}},\ and\
  \bibinfo {author} {\bibfnamefont {Y.}~\bibnamefont {Deng}},\ }\href
  {https://doi.org/10.1103/PhysRevE.107.044103} {\bibfield  {journal} {\bibinfo
   {journal} {Phys. Rev. E}\ }\textbf {\bibinfo {volume} {107}},\ \bibinfo
  {pages} {044103} (\bibinfo {year} {2023})}\BibitemShut {NoStop}%
\bibitem [{\citenamefont {Fang}\ \emph {et~al.}(2022)\citenamefont {Fang},
  \citenamefont {Zhou},\ and\ \citenamefont {Deng}}]{HighDFK2}%
  \BibitemOpen
  \bibfield  {author} {\bibinfo {author} {\bibfnamefont {S.}~\bibnamefont
  {Fang}}, \bibinfo {author} {\bibfnamefont {Z.}~\bibnamefont {Zhou}},\ and\
  \bibinfo {author} {\bibfnamefont {Y.}~\bibnamefont {Deng}},\ }\href
  {https://doi.org/10.1088/0256-307X/39/8/080502} {\bibfield  {journal}
  {\bibinfo  {journal} {Chin. Phys. Lett.}\ }\textbf {\bibinfo {volume} {39}},\
  \bibinfo {pages} {080502} (\bibinfo {year} {2022})}\BibitemShut {NoStop}%
\bibitem [{\citenamefont {Fang}\ \emph
  {et~al.}(2021{\natexlab{b}})\citenamefont {Fang}, \citenamefont {Zhou},\ and\
  \citenamefont {Deng}}]{FKCG2021}%
  \BibitemOpen
  \bibfield  {author} {\bibinfo {author} {\bibfnamefont {S.}~\bibnamefont
  {Fang}}, \bibinfo {author} {\bibfnamefont {Z.}~\bibnamefont {Zhou}},\ and\
  \bibinfo {author} {\bibfnamefont {Y.}~\bibnamefont {Deng}},\ }\href
  {https://doi.org/10.1103/PhysRevE.103.012102} {\bibfield  {journal} {\bibinfo
   {journal} {Phys. Rev. E}\ }\textbf {\bibinfo {volume} {103}},\ \bibinfo
  {pages} {012102} (\bibinfo {year} {2021}{\natexlab{b}})}\BibitemShut
  {NoStop}%
\bibitem [{\citenamefont {Li}\ \emph {et~al.}(2023)\citenamefont {Li},
  \citenamefont {Zhou}, \citenamefont {Fang},\ and\ \citenamefont
  {Deng}}]{LoopCG2023}%
  \BibitemOpen
  \bibfield  {author} {\bibinfo {author} {\bibfnamefont {Z.}~\bibnamefont
  {Li}}, \bibinfo {author} {\bibfnamefont {Z.}~\bibnamefont {Zhou}}, \bibinfo
  {author} {\bibfnamefont {S.}~\bibnamefont {Fang}},\ and\ \bibinfo {author}
  {\bibfnamefont {Y.}~\bibnamefont {Deng}},\ }\href
  {https://doi.org/10.1103/PhysRevE.108.024129} {\bibfield  {journal} {\bibinfo
   {journal} {Phys. Rev. E}\ }\textbf {\bibinfo {volume} {108}},\ \bibinfo
  {pages} {024129} (\bibinfo {year} {2023})}\BibitemShut {NoStop}%
\bibitem [{\citenamefont {Xiao}\ \emph {et~al.}(2024)\citenamefont {Xiao},
  \citenamefont {Li}, \citenamefont {Zhou}, \citenamefont {Fang},\ and\
  \citenamefont {Deng}}]{xiao2023finitesize}%
  \BibitemOpen
  \bibfield  {author} {\bibinfo {author} {\bibfnamefont {T.}~\bibnamefont
  {Xiao}}, \bibinfo {author} {\bibfnamefont {Z.}~\bibnamefont {Li}}, \bibinfo
  {author} {\bibfnamefont {Z.}~\bibnamefont {Zhou}}, \bibinfo {author}
  {\bibfnamefont {S.}~\bibnamefont {Fang}},\ and\ \bibinfo {author}
  {\bibfnamefont {Y.}~\bibnamefont {Deng}},\ }\href
  {https://doi.org/10.1103/PhysRevE.109.034125} {\bibfield  {journal} {\bibinfo
   {journal} {Phys. Rev. E}\ }\textbf {\bibinfo {volume} {109}},\ \bibinfo
  {pages} {034125} (\bibinfo {year} {2024})}\BibitemShut {NoStop}%
\bibitem [{\citenamefont {Swendsen}\ and\ \citenamefont
  {Wang}(1987)}]{SWalgorithm}%
  \BibitemOpen
  \bibfield  {author} {\bibinfo {author} {\bibfnamefont {R.~H.}\ \bibnamefont
  {Swendsen}}\ and\ \bibinfo {author} {\bibfnamefont {J.-S.}\ \bibnamefont
  {Wang}},\ }\href {https://doi.org/10.1103/PhysRevLett.58.86} {\bibfield
  {journal} {\bibinfo  {journal} {Phys. Rev. Lett.}\ }\textbf {\bibinfo
  {volume} {58}},\ \bibinfo {pages} {86} (\bibinfo {year} {1987})}\BibitemShut
  {NoStop}%
\bibitem [{\citenamefont {Wolff}(1989)}]{WolffAlg}%
  \BibitemOpen
  \bibfield  {author} {\bibinfo {author} {\bibfnamefont {U.}~\bibnamefont
  {Wolff}},\ }\href {https://doi.org/10.1103/PhysRevLett.62.361} {\bibfield
  {journal} {\bibinfo  {journal} {Phys. Rev. Lett.}\ }\textbf {\bibinfo
  {volume} {62}},\ \bibinfo {pages} {361} (\bibinfo {year} {1989})}\BibitemShut
  {NoStop}%
\bibitem [{\citenamefont {Wang}\ \emph {et~al.}(2002)\citenamefont {Wang},
  \citenamefont {Kozan},\ and\ \citenamefont {Swendsen}}]{PhysRevE.66.057101}%
  \BibitemOpen
  \bibfield  {author} {\bibinfo {author} {\bibfnamefont {J.-S.}\ \bibnamefont
  {Wang}}, \bibinfo {author} {\bibfnamefont {O.}~\bibnamefont {Kozan}},\ and\
  \bibinfo {author} {\bibfnamefont {R.~H.}\ \bibnamefont {Swendsen}},\ }\href
  {https://doi.org/10.1103/PhysRevE.66.057101} {\bibfield  {journal} {\bibinfo
  {journal} {Phys. Rev. E}\ }\textbf {\bibinfo {volume} {66}},\ \bibinfo
  {pages} {057101} (\bibinfo {year} {2002})}\BibitemShut {NoStop}%
\bibitem [{\citenamefont {Prokof'ev}\ and\ \citenamefont
  {Svistunov}(2001{\natexlab{a}})}]{WormAlg}%
  \BibitemOpen
  \bibfield  {author} {\bibinfo {author} {\bibfnamefont {N.}~\bibnamefont
  {Prokof'ev}}\ and\ \bibinfo {author} {\bibfnamefont {B.}~\bibnamefont
  {Svistunov}},\ }\href {https://doi.org/10.1103/PhysRevLett.87.160601}
  {\bibfield  {journal} {\bibinfo  {journal} {Phys. Rev. Lett.}\ }\textbf
  {\bibinfo {volume} {87}},\ \bibinfo {pages} {160601} (\bibinfo {year}
  {2001}{\natexlab{a}})}\BibitemShut {NoStop}%
\bibitem [{\citenamefont {El\ifmmode~\mbox{\c{c}}\else \c{c}\fi{}i}\ \emph
  {et~al.}(2018)\citenamefont {El\ifmmode~\mbox{\c{c}}\else \c{c}\fi{}i},
  \citenamefont {Grimm}, \citenamefont {Ding}, \citenamefont {Nasrawi},
  \citenamefont {Garoni},\ and\ \citenamefont {Deng}}]{LiftedWorm}%
  \BibitemOpen
  \bibfield  {author} {\bibinfo {author} {\bibfnamefont {E.~M.}\ \bibnamefont
  {El\ifmmode~\mbox{\c{c}}\else \c{c}\fi{}i}}, \bibinfo {author} {\bibfnamefont
  {J.}~\bibnamefont {Grimm}}, \bibinfo {author} {\bibfnamefont
  {L.}~\bibnamefont {Ding}}, \bibinfo {author} {\bibfnamefont {A.}~\bibnamefont
  {Nasrawi}}, \bibinfo {author} {\bibfnamefont {T.~M.}\ \bibnamefont
  {Garoni}},\ and\ \bibinfo {author} {\bibfnamefont {Y.}~\bibnamefont {Deng}},\
  }\href {https://doi.org/10.1103/PhysRevE.97.042126} {\bibfield  {journal}
  {\bibinfo  {journal} {Phys. Rev. E}\ }\textbf {\bibinfo {volume} {97}},\
  \bibinfo {pages} {042126} (\bibinfo {year} {2018})}\BibitemShut {NoStop}%
\bibitem [{\citenamefont {Shchur}\ and\ \citenamefont
  {Bl\"ote}(1997)}]{PseudoRN}%
  \BibitemOpen
  \bibfield  {author} {\bibinfo {author} {\bibfnamefont {L.~N.}\ \bibnamefont
  {Shchur}}\ and\ \bibinfo {author} {\bibfnamefont {H.~W.~J.}\ \bibnamefont
  {Bl\"ote}},\ }\href {https://doi.org/10.1103/PhysRevE.55.R4905} {\bibfield
  {journal} {\bibinfo  {journal} {Phys. Rev. E}\ }\textbf {\bibinfo {volume}
  {55}},\ \bibinfo {pages} {R4905} (\bibinfo {year} {1997})}\BibitemShut
  {NoStop}%
\bibitem [{\citenamefont {Edwards}\ and\ \citenamefont
  {Sokal}(1988)}]{PhysRevD.38.2009}%
  \BibitemOpen
  \bibfield  {author} {\bibinfo {author} {\bibfnamefont {R.~G.}\ \bibnamefont
  {Edwards}}\ and\ \bibinfo {author} {\bibfnamefont {A.~D.}\ \bibnamefont
  {Sokal}},\ }\href {https://doi.org/10.1103/PhysRevD.38.2009} {\bibfield
  {journal} {\bibinfo  {journal} {Phys. Rev. D}\ }\textbf {\bibinfo {volume}
  {38}},\ \bibinfo {pages} {2009} (\bibinfo {year} {1988})}\BibitemShut
  {NoStop}%
\bibitem [{\citenamefont {Grimmett}(2006)}]{grimmett2006random}%
  \BibitemOpen
  \bibfield  {author} {\bibinfo {author} {\bibfnamefont {G.}~\bibnamefont
  {Grimmett}},\ }\href@noop {} {\emph {\bibinfo {title} {The random-cluster
  model}}},\ Vol.\ \bibinfo {volume} {333}\ (\bibinfo  {publisher} {Springer},\
  \bibinfo {year} {2006})\BibitemShut {NoStop}%
\bibitem [{\citenamefont {Prokof'ev}\ and\ \citenamefont
  {Svistunov}(2001{\natexlab{b}})}]{PhysRevLett.87.160601}%
  \BibitemOpen
  \bibfield  {author} {\bibinfo {author} {\bibfnamefont {N.}~\bibnamefont
  {Prokof'ev}}\ and\ \bibinfo {author} {\bibfnamefont {B.}~\bibnamefont
  {Svistunov}},\ }\href {https://doi.org/10.1103/PhysRevLett.87.160601}
  {\bibfield  {journal} {\bibinfo  {journal} {Phys. Rev. Lett.}\ }\textbf
  {\bibinfo {volume} {87}},\ \bibinfo {pages} {160601} (\bibinfo {year}
  {2001}{\natexlab{b}})}\BibitemShut {NoStop}%
\bibitem [{\citenamefont {Zinn-Justin}(2021)}]{zinn2021quantum}%
  \BibitemOpen
  \bibfield  {author} {\bibinfo {author} {\bibfnamefont {J.}~\bibnamefont
  {Zinn-Justin}},\ }\href@noop {} {\emph {\bibinfo {title} {Quantum field
  theory and critical phenomena}}},\ Vol.\ \bibinfo {volume} {171}\ (\bibinfo
  {publisher} {Oxford university press},\ \bibinfo {year} {2021})\BibitemShut
  {NoStop}%
\bibitem [{\citenamefont {Higgs}(1964)}]{Higgs1964}%
  \BibitemOpen
  \bibfield  {author} {\bibinfo {author} {\bibfnamefont {P.~W.}\ \bibnamefont
  {Higgs}},\ }\href {https://doi.org/10.1103/PhysRevLett.13.508} {\bibfield
  {journal} {\bibinfo  {journal} {Phys. Rev. Lett.}\ }\textbf {\bibinfo
  {volume} {13}},\ \bibinfo {pages} {508} (\bibinfo {year} {1964})}\BibitemShut
  {NoStop}%
\bibitem [{\citenamefont {Pekker}\ and\ \citenamefont
  {Varma}(2015)}]{higgsmodeinCMP}%
  \BibitemOpen
  \bibfield  {author} {\bibinfo {author} {\bibfnamefont {D.}~\bibnamefont
  {Pekker}}\ and\ \bibinfo {author} {\bibfnamefont {C.}~\bibnamefont {Varma}},\
  }\href {https://doi.org/10.1146/annurev-conmatphys-031214-014350} {\bibfield
  {journal} {\bibinfo  {journal} {Annual Review of Condensed Matter Physics}\
  }\textbf {\bibinfo {volume} {6}},\ \bibinfo {pages} {269} (\bibinfo {year}
  {2015})}\BibitemShut {NoStop}%
\end{thebibliography}%

\end{document}